\documentclass{aa}  

\usepackage{graphicx}
\usepackage{color}

\usepackage{txfonts}
\begin{document} 

   \title{Dips in the sub-TeV gamma-ray light curves from central parts of galaxies 
due to transiting luminous stars}
   \titlerunning{Dips in gamma-ray light curves from galaxies}

   \author{W. Bednarek, J. Sitarek \& M. Ulatowski
          }

   \institute{Department of Astrophysics, Faculty of Physics and Applied Informatics,  University of Lodz,
              ul. Pomorska 149/153, 90-236 Lodz, Poland\\
              \email{wlodzimierz.bednarek@uni.lodz.pl; julian.sitarek@uni.lodz.pl; gamikolajulatowski@gmail.com}
             }

   \date{Received ... 2024; accepted ... , 2025}
 
  \abstract
   {The GeV-TeV $\gamma$-ray emission is observed from the direction of the source Sgr A$^\star$, which is identified with the Super-massive Black Hole (SMBH) in the centre of our Galaxy. 
According to some models this $\gamma$-ray emission might originate in the very compact, central region identified with the direct surrounding of the SMBH.   
Sgr A$^\star$ is surrounded by a massive nuclear star cluster, composed of the so-called S stars, also including OB-type stars.   
Occasionally these stars might pass close to the line of sight of the observer, resulting in partial absorption of the sub-TeV $\gamma$-ray emission.
}
   {
   We investigate the conditions at which an absorption feature appears in the $\gamma$-ray light curves from the Galactic Centre or nuclei of other galaxies containing SMBHs. 
   The detection of such features would allow to obtain constraints on the emission site of $\gamma$ rays in active galaxies. 

}
   {
    We calculate the optical depths for $\gamma$ rays in the radiation of individual massive stars, or from the whole population of stars for different parameters of the star cluster.} 
   {
We show that the observer with a line of sight close to the orbital plane of the star can register a $\gamma$-ray absorption dip lasting from a fraction of a day up to a few tens of days.   
The combined effect of the bulk absorption on the whole population of stars instead can produce a flickering of the observed emission of a red-noise type in the power spectrum of the emission. 
   }
   {Predicted absorption features in the sub-TeV $\gamma$-ray light curves from galaxies with SMBHs should be easily detectable by the Large-Sized Telescopes of the Cherenkov Telescope Array Observatory. The discovery of such absorption features provides a unique indication
that the $\gamma$-ray production is occurring in a compact region, close to the horizon of the SMBH.}


   \maketitle
%

\section{Introduction}

The central parsec around the Super-massive Black Hole (SMBH) within the Galactic Centre (GC) (plausibly associated with the Sgr A$^\star$ source)
contains about 10 million stars \citep{1995ApJ...447..L95}. In fact, already about a hundred luminous OB and Wolf-Rayet type star are observed \citep{1991ApJ...382..L19} but much more are expected from the comparison with other galaxies and also the content of Galactic star clusters. \citet{2010ApJ...708..834} reports the presence of  $N_\star = 177$ O/WR/B stars  at a typical distance of 0.03~pc from Sgr A$^\star$. 
Two extreme examples of WR type stars close to Sgr A$^\star$ (WR 102ka and WR 102c) have luminosities of the order of a few million $L_\odot$ \citep{2008A&A...486..971} and nineteen of them have luminosities close to $10^6$~L$_\odot$ (with L$_\odot$ being the luminosity of the Sun) and 88 are of Wolf-Rayet type \citep{2010ApJ...725..188}. 

Moreover, the so-called S stars have been discovered around Sgr A$^\star$ on a very tight, randomly oriented orbits within the range of distances 0.01 to 0.1~pc from the SMBH \citep{2005ApJ...620..744}. 
One of these extreme stars (named S2), which is massive and luminous, has eccentricity $e = 0.88429\pm 0.00006$, and the pericenter passage $1.8\times 10^{16}$~cm \citep{2019A&A...625L..10G}. 
These stars were probably formed in an accretion disk around SMBH a few million years ago.     
The simulations of the evolution of the central star cluster show that as a result of stellar collisions 
about $\sim$100 massive ($>10$~M$_\odot$, with M$_\odot$ being the solar mass) stars 
can reside between 0.01 and 0.1~pc from the SMBH within Galactic Centre \citep{2023ApJ...955..30}.
Therefore, the close surrounding of the SMBH within our Galaxy is full of stars including luminous massive stars and red giants \citep{2010RMP...82..3121}. 
The observations of other massive stellar clusters in our Galaxy indicate that the numbers of luminous stars
around Sgr A$^\star$ can be even much larger. For example,
in the open cluster within our Galaxy Cyg OB2, with the total mass equal to $(4 - 10)\times 10^4$~M$_\odot$,  the number of luminous OB type stars is $2600\pm 400$ and O type stars is $120\pm 20$ O \citep{2000A&A...360..539K}. In the so-called super star cluster Westerlund 1, the mass in stars is $10^4$~M$_\odot$ and about $10^3$ O type and 28 WR type stars were discovered in this super cluster \citep{2024arXiv240811087A}.

The centre of our Galaxy, Sgr A$^\star$, is the source of non-thermal radiation in the whole range of the electromagnetic spectrum. This emission extends up to the TeV $\gamma$-ray energies  \citep[e.g.][]{2004A&A...425L..13A,2004ApJ...608L..97K,2004ApJ...606L.115T,2006ApJ...638L.101A,2024ApJ...submitted}. This TeV emission is expected to come (at least in significant amount) from Sgr A$^\star$ itself. The observations at lower energies with the \textit{Fermi}-LAT reports the point-like source centred directly on the Sgr A$^\star$ (e.g. \citealp{2021ApJ...918..30}). According to some models, the $\gamma$-ray emission might be produced
close to the horizon of the SMBH as a result of the acceleration of particles in the SMBH magnetosphere \citep{2007ApJ...671..85,2008A&A...479..L5,2011ApJ...730..123}, 
in a region with the size comparable to the SMBH horizon, i.e. $\sim 10^{12}$~cm for the SMBH mass equal to $(3.7\pm 0.2)\times 10^6$~M$_\odot$ \citep{2005ApJ...620..744}. Such a production mechanism provides an interesting explanation for the very short flares observed e.g. in blazar IC~310 \citep{2014Science...346..1080,2018Galax...6..122H}.
It is expected \citep{2005ApJ...619..306}, that in the case of Sgr A$^\star$ such TeV $\gamma$-ray emission might be able to escape from a relatively weak low-energy radiation field around SMBH.

If in fact sub-TeV $\gamma$-ray emission originates close to the vicinity of the SMBH, then massive stars in nuclear star cluster can cross from time to time close the line of sight of the SMBH by the observer on the Earth. Then, the sub-TeV $\gamma$-ray emission from the close vicinity of the SMBH is expected to be partially absorbed, forming the characteristic absorption spectral feature. 
Such absorption features, but in sub-TeV $\gamma$-ray light curves from jetted active galaxies, 
have been recently investigated \citep{2021MNRAS...503..2423B}.  

Note that nuclear star clusters are observed in most galaxies. They have masses of $10^6-10^7$~M$_\odot$ \citep{2006ApJ...649.692} within the volume of 2-5~pc \citep{2004AJ...127..105}. They mainly surround lower-mass SMBHs, i.e. below $10^{10}$~M$_\odot$ \citep{2009MNRAS...397..2148}. Therefore, we expect that similar transient absorption features in the sub-TeV $\gamma$-ray spectra from galaxies containing intermediate SMBHs should also be observed.  For example, one of the closest galaxies, NGC~4945 (distance $3.8\pm 0.3$~Mpc, \citealp{2007AJ...133..504}), contains a SMBH that is similar to the one in our Galaxy  (i.e. with the mass $M_{\rm SMBH} = 1.4\times 10^6$~M$_\odot$, \citealp{1997ApJL...481..L23}). 
Notably, GeV emission has been seen from NGC~4945 by \textit{Fermi}-LAT (Large Area Telescope) \citep{2010ApJS..188..405A, 2010A&A...524A..72L}. 
NGC~4945 is surrounded by several star clusters with the present size of 
2-3 pc (forming a super star cluster) that contain the total stellar mass of $1.1\times 10^7$~M$_\odot$ \citep{2020ApJ...903..50}. By scaling from the number of O type stars in Galactic star clusters we estimate that around the SMBH in NGC~4945 about $\sim 10^4$ O type stars should be present. It is expected that at some moment the super star cluster around NGC~4945 will collapse to the sizes observed in our Galaxy.

Strong radiation from massive stars (or their clusters) propagating within (or close) to the jets in active galaxies, can also provide a strong target for the production of $\gamma$ rays by relativistic leptons in the jet (e.g. \citealp{1997MNRAS...287L..9B,2016MNRAS...463L..26B}). On the other hand, if $\gamma$ rays are produced in the inner jet launched from the SMBH, they can be absorbed in the radiation of luminous stars which passes through the jet close to the observer's line of sight \citep{2021MNRAS...503..2423B}. Here we propose that absorption effects in the sub-TeV $\gamma$-ray spectra can also be observed in the case of normal galaxies whose SMBHs are surrounded by central star clusters, provided that they produce sub-TeV $\gamma$ rays close to the SMBHs with mild masses (i.e. below $10^9$~M$_\odot$). We show that such absorption features can last from a fraction of a day up to a few tens of days. They have a chance to be discovered with future telescopes such as Large-Sized Telescopes (LST, \citealp{2023ApJ...956...80A}) of the future Cherenkov Telescope Array Observatory (CTAO, \citealp{2022icrc.confE...5Z}). Their discovery would provide smoking gun 
arguments for the origin of $\gamma$-ray emission in the direct vicinity of the SMBHs in centres of galaxies.

\section{Model for the dip formation}

\begin{figure*}[t]
\includegraphics[width=0.4\textwidth]{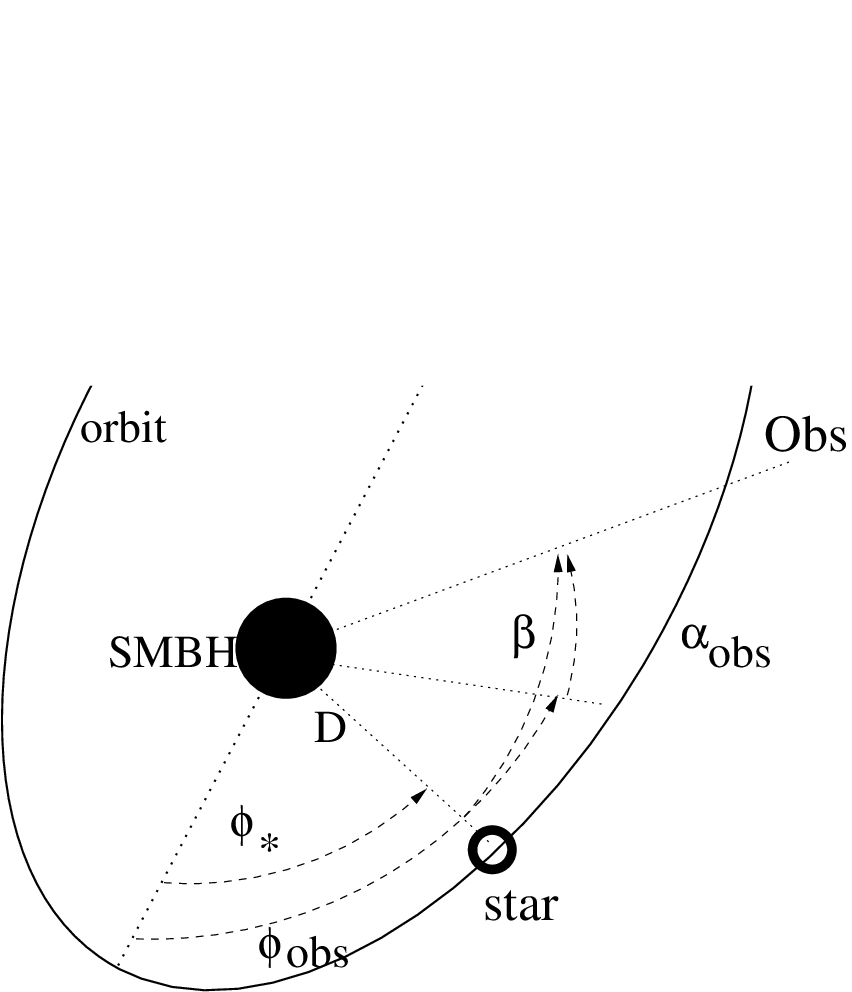}
\includegraphics[width=0.5\textwidth]{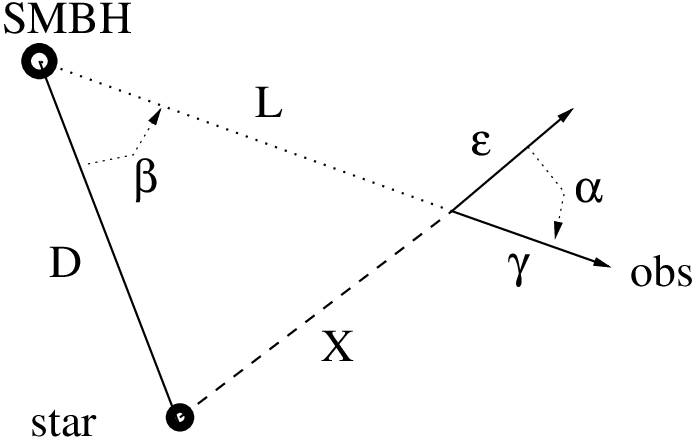}
\caption{
On the left: Schematic representation of a massive star moving on an elliptic orbit around the SMBH. The phase of the star is marked by $\Phi_{\star}$ and its distance from the SMBH is $D$. The observer (obs)
is located at the angle $\alpha_{\rm obs}$ in respect to the plane of the orbit and its phase is 
$\Phi_{\rm obs}$. $\beta$ is the angle between the direction defined by the SMBH and the star and the direction defined by the SMBH and the observer.
On the right: The geometry of the interaction between the $\gamma$-ray photon ($\gamma$) and the soft photon 
($\varepsilon$) from the star. 
$L$ is the propagation distance of the $\gamma$ ray and $\alpha$ is the incident angle of the colliding photons.}
\label{fig1}
\end{figure*}

Some models of the $\gamma$-ray production in active galaxies argue that they are produced in a very compact region comparable to the size of the horizon of the SMBH and at large angles to the direction of relativistic jets (see references in the Introduction). 
It should happen from time to time that a massive star from the nuclear star cluster transits between the $\gamma$-ray source and the distant observer (see Fig.~\ref{fig1} on the left).  
Then, radiation field of early type stars, of the O, B and WR types, can form a strong enough target for the $\gamma$ rays (typically with the sub-TeV energies). As a result, sub-TeV $\gamma$-ray emission from the vicinity of the SMBH can become partially absorbed. 

The observer should see the characteristic absorption dip in the $\gamma$-ray light curve observed from the nucleus of this galaxy. A typical time scale of the reduced emission depends on the velocity of the luminous star and its parameters (the radius, the surface temperature). In a simple model of circular orbits, the velocity of the star, $v_\star$, depends on its distance from the SMBH, $D = 10^{-2}D_{-2}$~pc, and the SMBH mass, $M_{\rm SMBH} = 10^8M_8$~M$_\odot$, as:
\begin{eqnarray}
v_\star = (G M_{\rm SMBH}/D)^{1/2}\approx 7\times 10^8{M_8}^{1/2}D_{-2}^{-1/2}~~~
{\rm cm~s^{-1}}.
\label{eq1}
\end{eqnarray}
\noindent
The density of photons from the massive star can be
estimated from 
\begin{equation}
n_{\rm ph} = {\frac{L_\star}{3k_{\rm B} T_\star4 \pi X^2c}} \approx  
1.2\times 10^{14}{\frac{T_{4.5}^3}{r^2}}~\mathrm{ph.~cm}^{-3},
\label{eq2}    
\end{equation}
where $c$ is the velocity of light, X is the distance of the $\gamma$-ray from the centre of the star, the surface temperature of the star is $T_\star = 3\times 10^4T_{4.5}$~K, 
its radius is $R_\star = 10^{12} R_{12}$~cm, its luminosity is $L_\star$, and $r=X/R_\star$. The mean free path for $\gamma$ rays can be estimated from $\lambda_{\gamma\gamma} = (n_{\rm ph}\sigma_{\gamma\gamma})^{-1}$, 
where $\sigma_{\gamma\gamma}\approx \sigma_{\rm T}/3$, is the cross section for $e^\pm$ pair production in 
collisions of two photons and the Thomson cross section is $\sigma_{\rm T} = 6.65\times 10^{-25}$~cm$^2$. We can
introduce the condition for the efficient absorption of $\gamma$ rays, when the mean free path becomes comparable to the propagation distance in the stellar field, i.e. 
$\lambda_{\gamma\gamma} \approx X$. 
This condition allows us to estimate the region around the star, dubbed the radius of the $\gamma$-sphere, in which $\gamma$ rays are absorbed. For an O type star, the radius of the $\gamma$-sphere, $R_{\gamma}$, expressed in units of the stellar radius:
\begin{eqnarray}
r_{\gamma}^{\rm O} = R_{\gamma}/R_\star \approx 27 T_{4.5}^3R_{12}.
\label{eq3}
\end{eqnarray}
Then, the duration of the eclipse of the central $\gamma$-ray source by the soft radiation from the moving luminous star (referred to in the following as the dip) is: 
\begin{eqnarray} 
T_{\rm dip}\approx 2R_{\gamma}v_\star^{-1}\approx 
0.97T_{4.5}^3R_{12}D_{-2}^{1/2}M_8^{-1/2}~~~{\rm days}. 
\end{eqnarray}
\noindent
For the SMBH in the GC ($M_{\rm SMBH} = 4\times 10^6$~M$_\odot$), the dip in the $\gamma$-ray light curve should be observed during a few days.
However, for the typical mass of the SMBH in the centre of active galaxies
($M_{\rm SMBH} = 10^8$~M$_\odot$), it should last on a sub-day time scale assuming that the star crosses the
direction to the observer at a typical distance of $D = 10^{-2}$~pc.
Note however, that in order to detect significant change in the sub-TeV $\gamma$-ray emission, the dimension of the $\gamma$-sphere has to be clearly larger than the dimension of the SMBH horizon, i.e. 
$R_{\rm Sch} = 29 \times 10^{12} M_8$~cm.

In the case of very luminous stars of the WR type, discovered within Central Cluster \citep{2010ApJ...725..188}, the duration of the dip in the light curve can be even a few times larger than estimated above.

\section{The effect of stellar transition on the $\gamma$-ray propagation}\label{sec:transit}  

We calculate the effects of the absorption of the $\gamma$-ray emission from a core of an active galaxy assuming that a massive star moves on an elliptic orbit around the SMBH.
The distance of the star from the SMBH, in the polar reference frame, is described by 
$D = a (1 - e^2)/(1 + e \cos{\phi_{\star}})$, where $e$ is the ellipticity. 
The location of the observer is determined by its phase, $\phi_{\rm obs}$, and the inclination of the orbital plane to the observer's line of sight, $\alpha_{\rm obs}$ (see Fig.~1).
The phase of the star is measured from the semi-major axis of the stellar orbit $\phi_{\star}$ (see Fig.~1 on the left for details). 
In order to calculate the optical depth for the $\gamma$ rays produced close to the SMBH in the radiation of the star, we have to calculate the angle, $\beta$, between the direction towards the star and the observer as seen from the location of the SMBH. It is given by 
$\cos{\beta} = \cos{(\phi_{\rm obs}} - \phi_\star) \times \cos{\alpha_{\rm obs}}$ (Fig.~1).
The angle, $\beta$, allows us to determine the distance of the $\gamma$-ray photon from the centre of the star
for its known propagation distance, L, i.e. $X^2 = D^2 + L^2 - 2 D L \cos{\beta}$. 
Then, we obtain the incident angle, $\alpha$, between the $\gamma$-ray photon and the soft photon from the direction towards the centre of the star (see Fig.~1 on the right),
$\cos\alpha = 0.5(X^2 + L^2 - D^2)/(X L)$. 

Knowing the location of the star on its orbit around SMBH and the location of the observer, we can calculate the optical depths for the $\gamma$-ray photons, produced in the vicinity of the SMBH, from:
\begin{eqnarray} 
\tau_{\gamma\gamma}(E_\gamma) &=& \int_{0}^{L_{\rm max}} dL  \int_{\varepsilon_{\rm min}}^{\infty} d\varepsilon
\int_{\Omega}{\frac{dn_{\rm ph}(\varepsilon,\Omega)}{d\Omega d\varepsilon}} \nonumber\\
& &(1 - \cos\alpha)\sigma_{\gamma\gamma}(E_\gamma, \varepsilon, \cos\alpha)     d\Omega, 
\label{eq5}
\end{eqnarray}
\noindent
where $dn_{\rm ph}(\varepsilon,\Omega)/d\Omega d\varepsilon = (2/h^3c^3)(\varepsilon^2/(\exp{(\varepsilon/k_{\rm B}T_\star)} - 1)$ is the differential photon density (ph./GeV/cm$^3$/sr), 
$\Omega$ is the solid angle of the star as seen from the location of the $\gamma$-ray photon,  
$\sigma_{\gamma\gamma}(E_\gamma, \varepsilon, \cos\alpha)$ is the energy and angle dependent cross section for two photon collisions in which $e^\pm$ pair is produced \citep{lang1999}, 
$\varepsilon_{\rm min} = 2(m_{\rm e}c^2)^2/[E_\gamma(1 - \cos\alpha)]$ is the minimum energy of the soft photon which is able to create an $e^\pm$ pair in collisions with the $\gamma$-ray with energy $E_\gamma$ for the photon incident angle  $\alpha$, 
$L_{\rm max}$ is the maximum propagation distance for which the $e^\pm$ pair can be produced,
and $m_{\rm e}c^2$ is the electron rest energy, $h$ is the Planck constant, $k_{\rm B}$ is the Boltzmann constant. 

\begin{figure*}[t]
\includegraphics[width=0.265\textwidth, trim=0 0 42 0, clip]{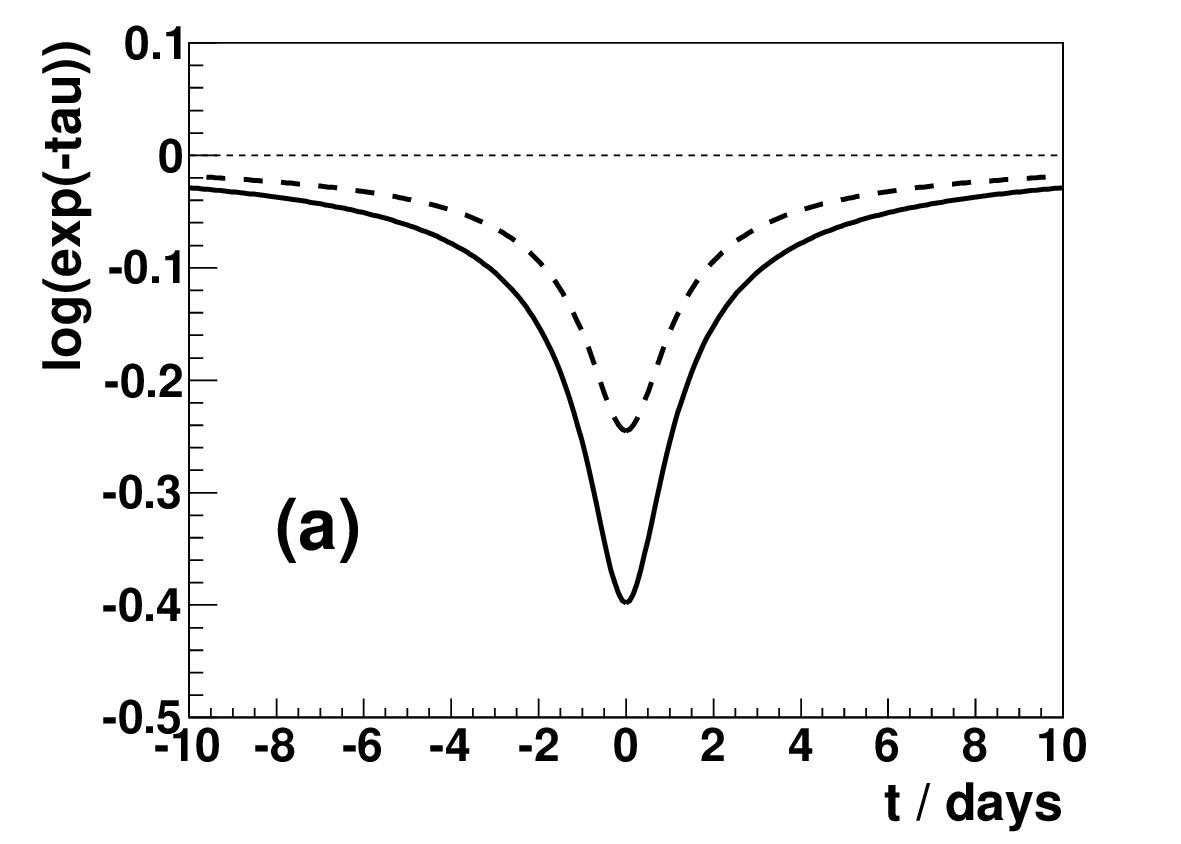}
\includegraphics[width=0.24\textwidth, trim=50 0 42 0, clip]{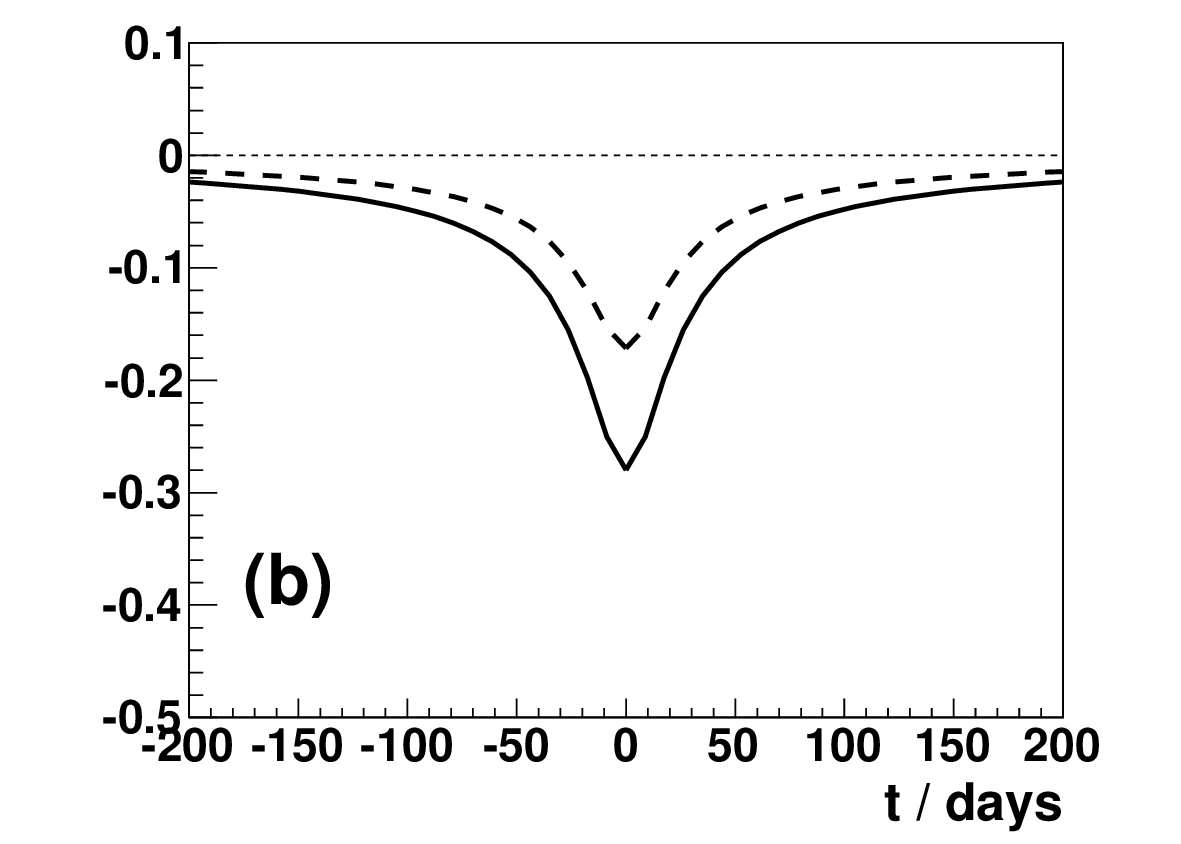}
\includegraphics[width=0.24\textwidth, trim=50 0 42 0, clip]{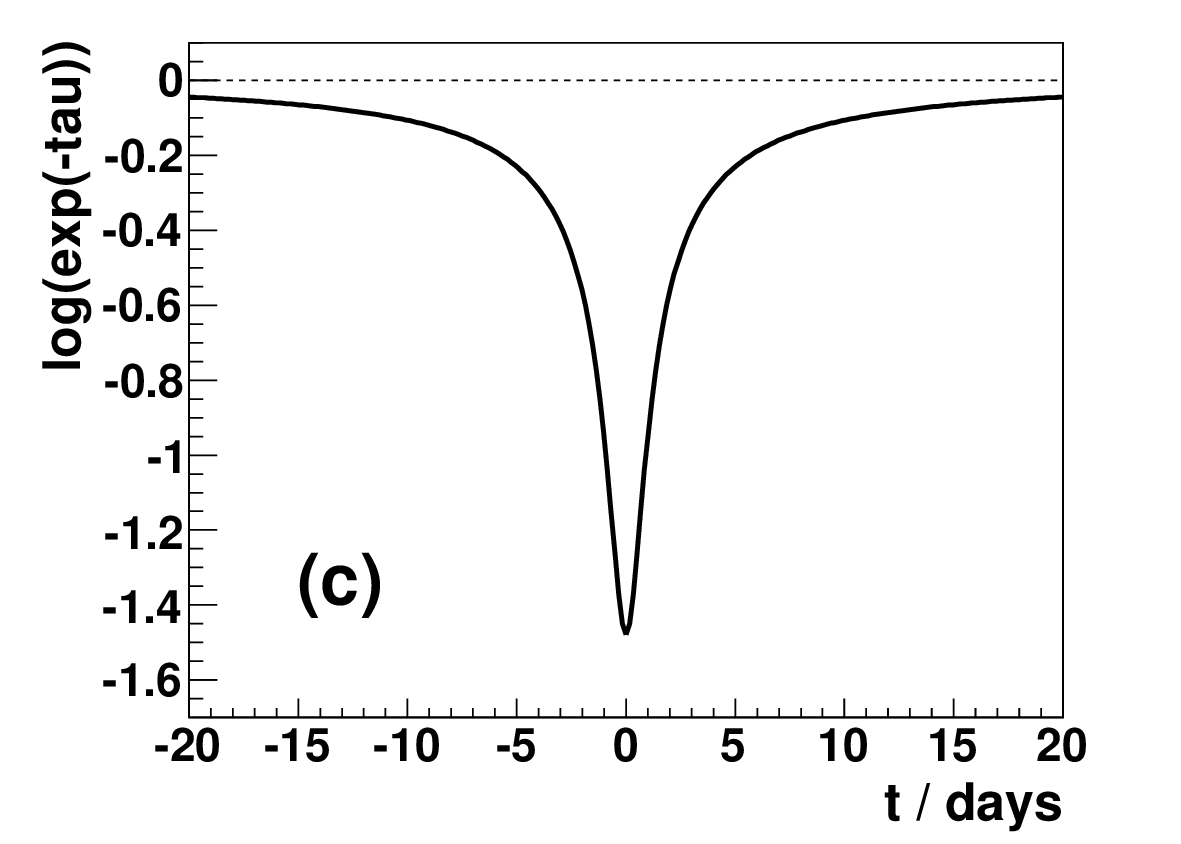}
\includegraphics[width=0.24\textwidth, trim=50 0 42 0, clip]{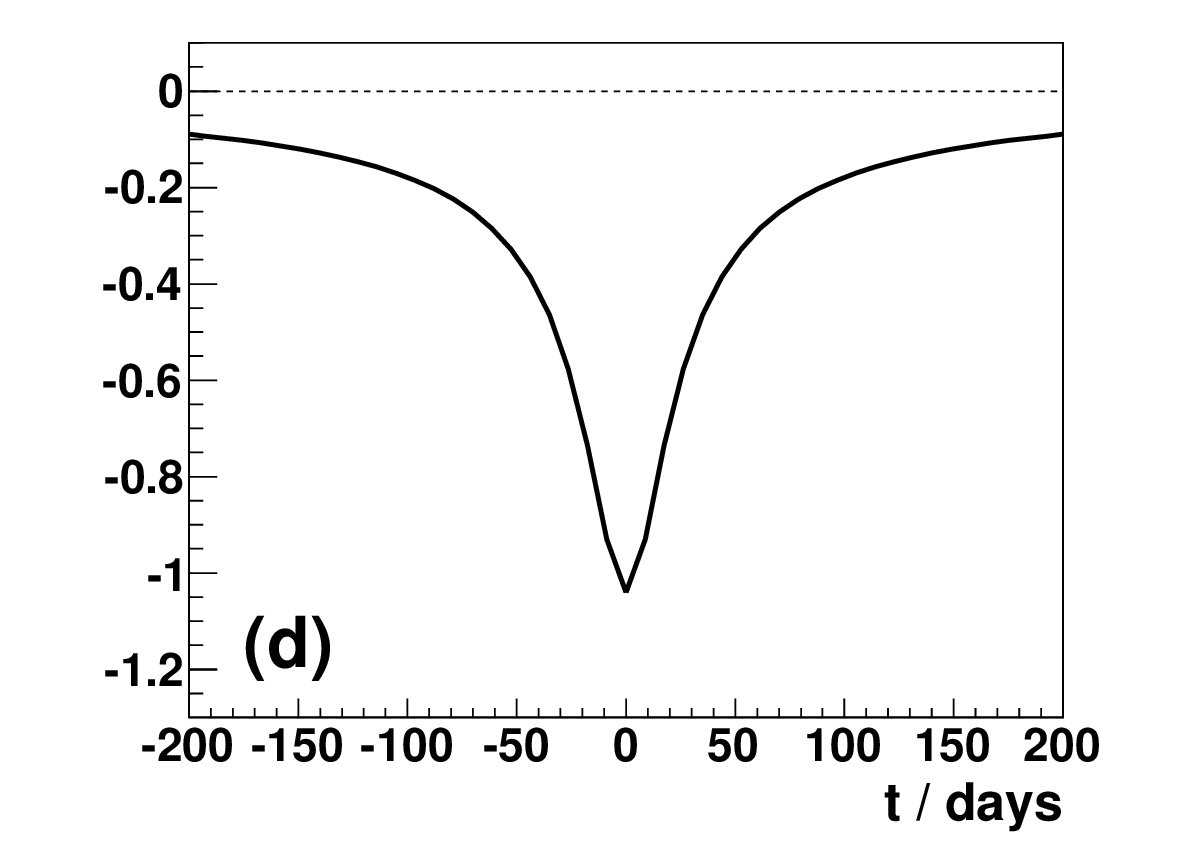}
\includegraphics[width=0.265\textwidth, trim=0 0 42 0, clip]{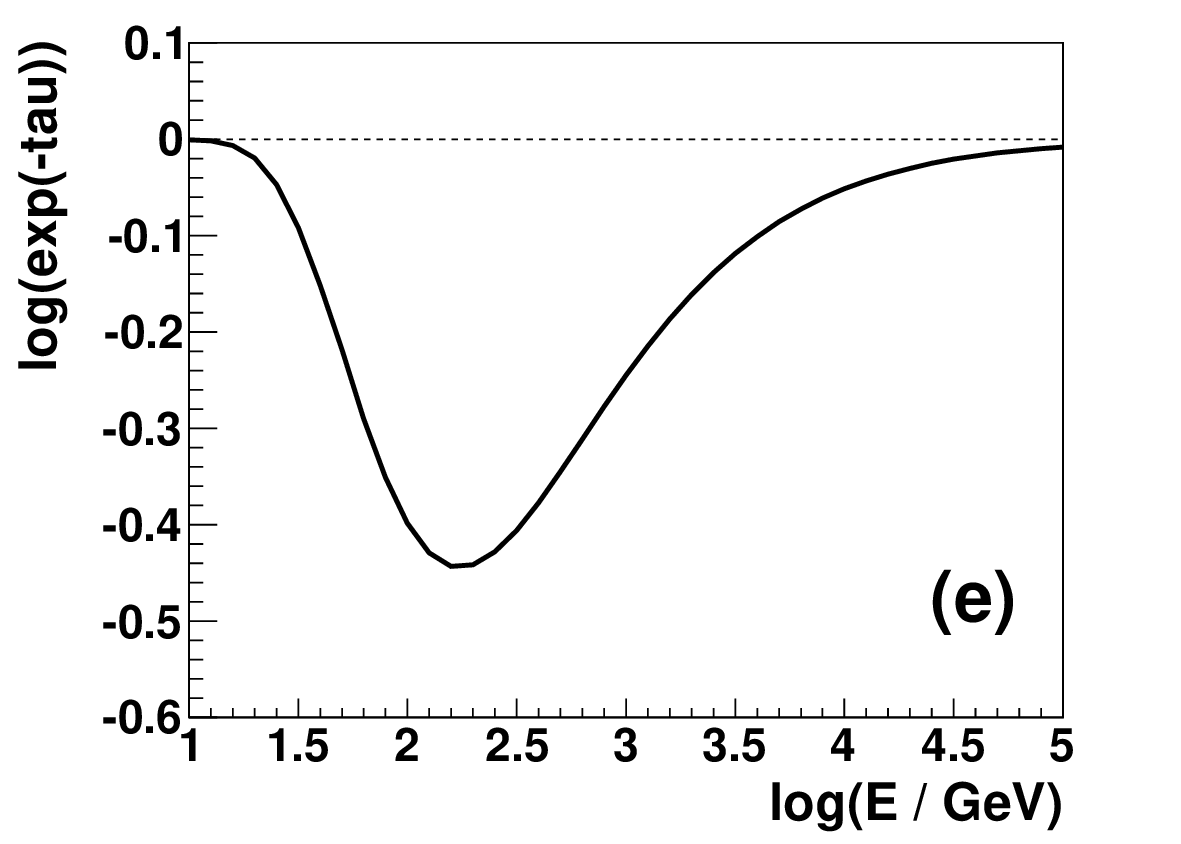}
\includegraphics[width=0.24\textwidth, trim=50 0 42 0, clip]{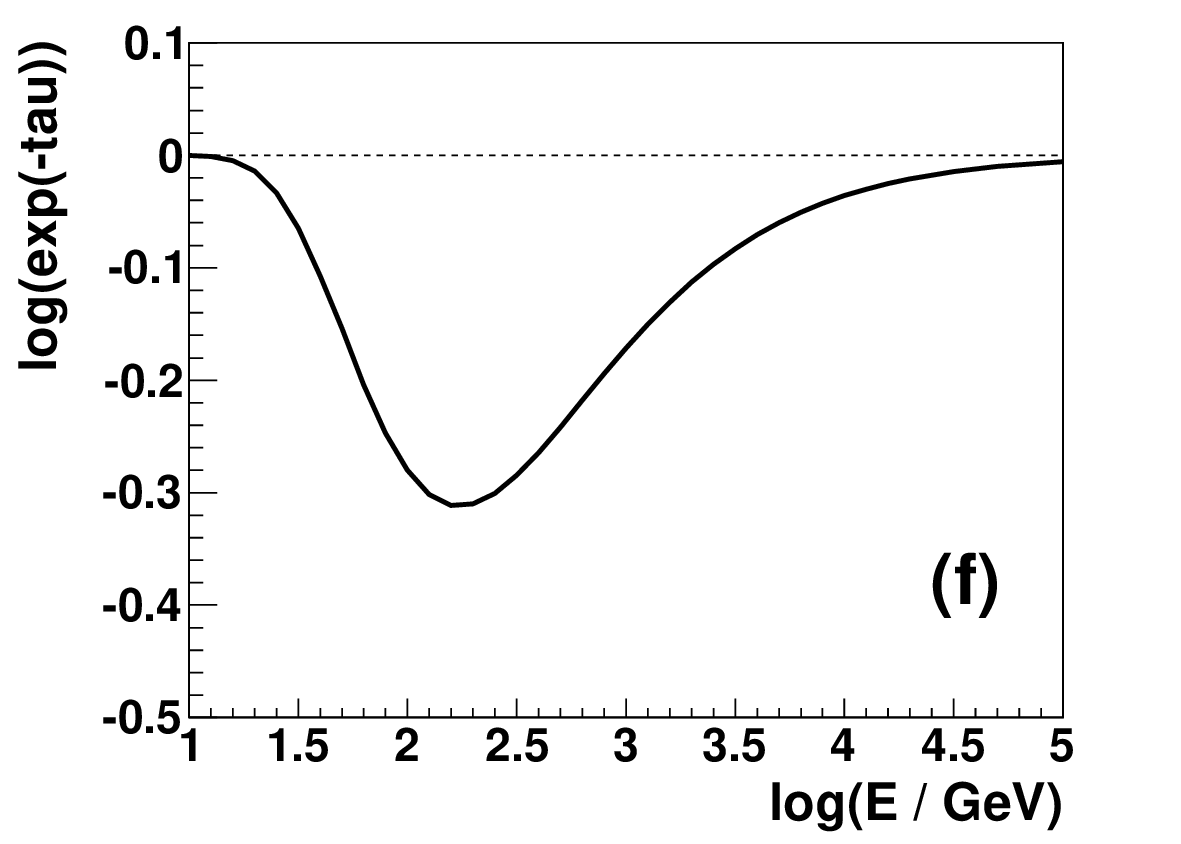}
\includegraphics[width=0.24\textwidth, trim=50 0 42 0, clip]{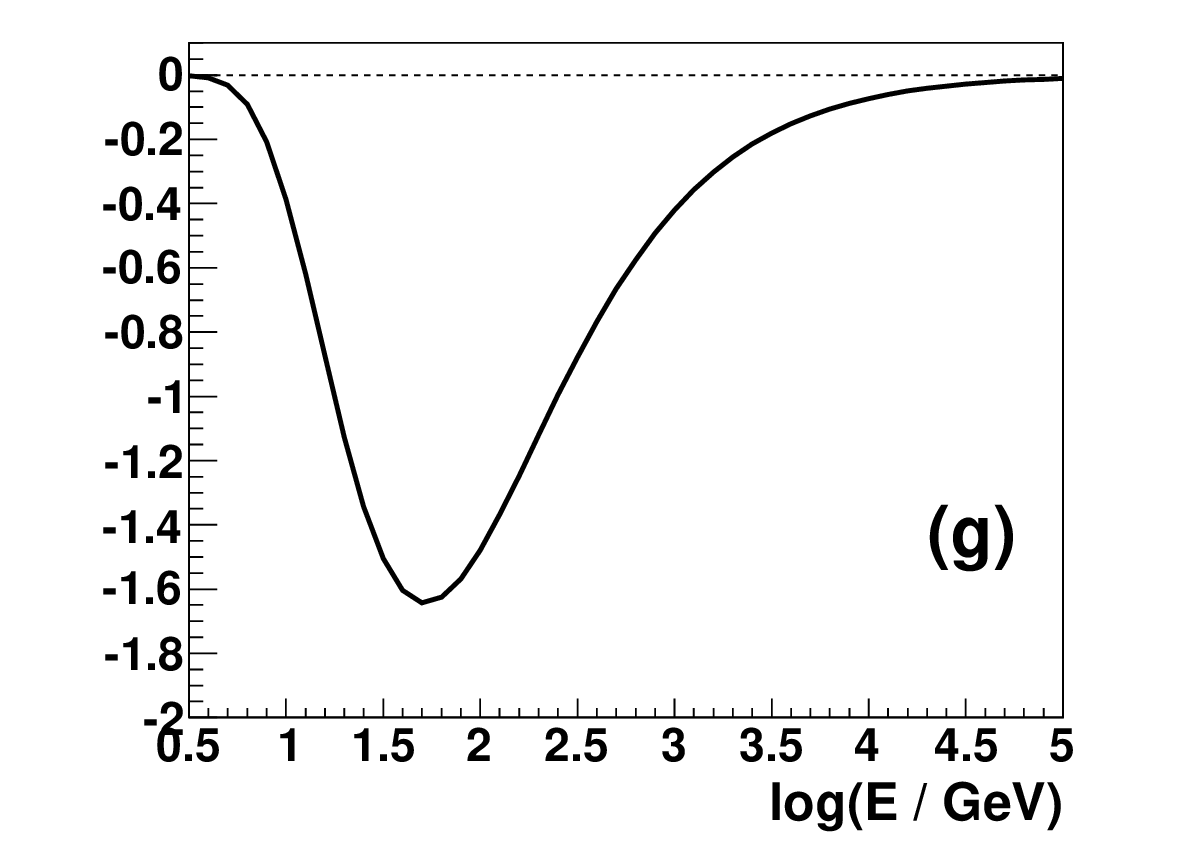}
\includegraphics[width=0.24\textwidth, trim=50 0 42 0, clip]{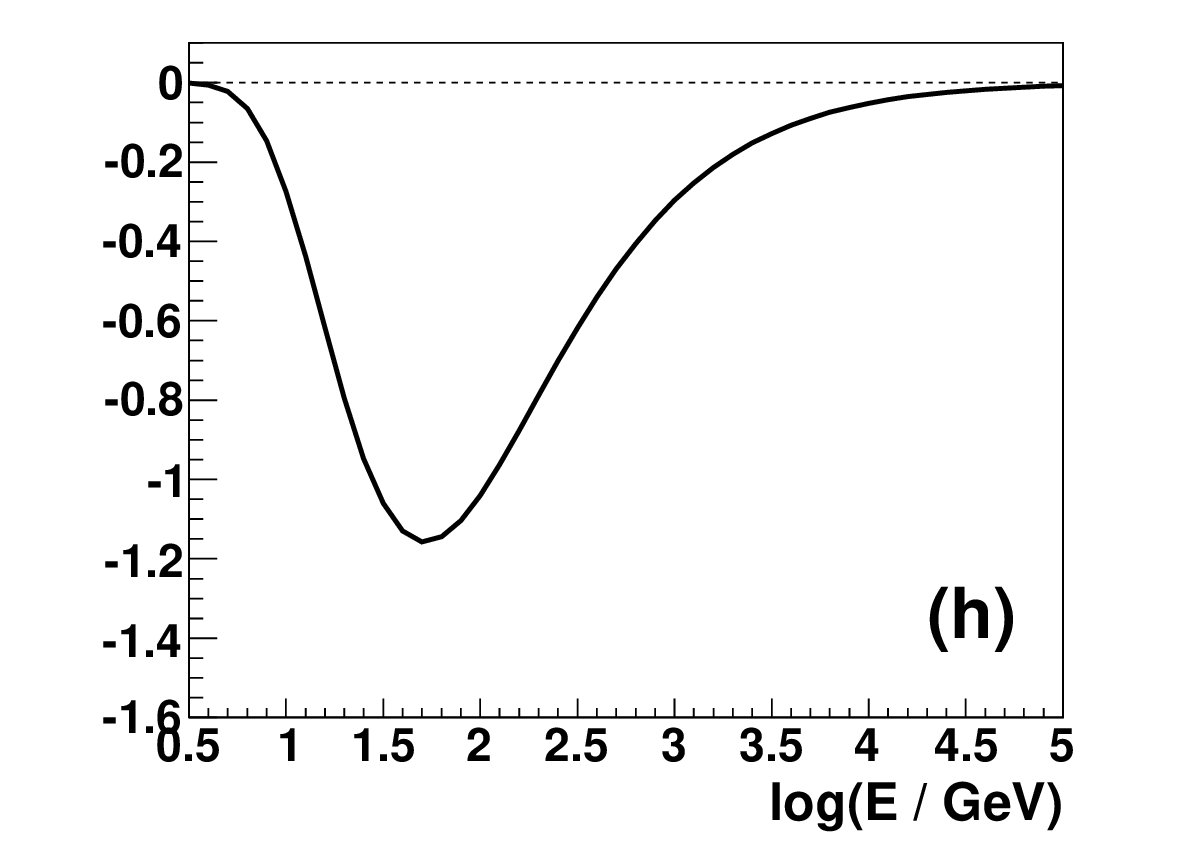}
\caption{Top figures: The reduction factor of the $\gamma$-ray emission at energy $E_\gamma = 100$~GeV (solid curve) and at 1 TeV (dashed curve), $R = \exp(-\tau)$, as a function of time measured from the perycenter of the star on its orbit around the SMBH with the mass  $4\times 10^6$~M$_\odot$. The orbit of the star is defined by the semi-major axis $a = 2\times 10^{16}$~cm and the eccentricity $e = 0.87$ (as observed in the case of the star S2 on the orbit around Sgr A$^\star$). In figures (a), (b), the star has the radius $R_\star = 10^{12}$~cm and the surface temperature $T_{\rm O} = 3\times 10^4$~K. The light curves are shown for the perycenter (figure a) and the apocenter (b) passages of this star. The light curves for the WR type star ($T_{\rm WR} = 10^5$~K and $R_{\rm WR} = 3\times 10^{11}$~cm) at the perycenter and apocenter are shown in figures (c) and (d), respectively. 
The observer is located at the angle $\alpha_{\rm obs} = 1^\circ$ in (a) and (c)) and $0.1^\circ$ in (b) and (d) in respect to the plane of the orbit  and at the phase $\Phi_{\rm obs} = 0^\circ$ measured from the perycenter or the apocenter of the stellar orbit. 
Bottom figures: The reduction factors, as a function of the $\gamma$-ray energy, for the parameters in the corresponding top panel.}
\label{fig2}
\end{figure*}

As an example, we consider a hypothetical luminous star, with the orbital parameters of the S2 star, moving within the central spherical star cluster around Sgr A$^\star$ between distances, pericenter $\sim 0.01$~pc and apocenter 
$\sim 0.1$~pc, but at small inclination angle to the line of sight. We consider the luminous star with our typical scaling parameters, i.e. $T_\star = 3\times 10^4$~K and $R_\star = 10^{12}$~cm. Note that these parameters are not so far from the parameters derived for the S2 star itself, i.e. surface temperature $T_{\rm S2} = 28513^{+2388}_{-2923}$~K and the radius $R_{\rm S2}/R_\odot = 5.53^{+1.77}_{-0.79}$ of the S2 star \citep{2017ApJ...847..120}.  
We obtain the dependence of the optical depths for the $\gamma$ rays on the polar angle of the star on its orbit around SMBH, the so-called true anomaly.
We obtain the eccentric anomaly by solving the equation relating it to the true anomaly.
Next, we solve the Kepler equation in order to obtain the dependence of the optical depth 
on time measured from the perycenter.
In Fig. \ref{fig2} we show how the reduction factors of the $\gamma$-ray emission, $R = \exp(-\tau_{\gamma\gamma})$, change with time around the perycenter  ($\phi_\star = 0^\circ$, Fig.~\ref{fig2}a), and apocenter ($\phi_\star = 180^\circ$ Fig.~\ref{fig2}b) in the case of considered star with the example parameters. In these cases the observer is located at the phase $\phi_{\rm obs} = 0^\circ$ and at a very small inclination angle equal to $\alpha_{\rm obs} = 1^\circ$, to the orbital plane of this star (see Fig.~\ref{fig2}). We show that for these parameters of the star and the observer, due to the absorption, the $\gamma$-ray flux can be reduced by almost an order of magnitude. This reduction occurs around the $\gamma$-ray energies $\sim 100$ GeV (see Figs. \ref{fig2}e,f). When the star is close to the apocenter then the reduction of the $\gamma$-ray emission becomes similar, but for much smaller inclination angle of the orbital plane of the star (close to $\alpha_{\rm obs} = 0.1^\circ$).
We also investigate the effects of $\gamma$-ray absorption in the case of the transit of a very luminous star of the WR type (T$_{\rm WR} = 10^5$~K and $R_{\rm WR} = 3\times 10^{11}$~cm). In such a case, the absorption effect on the $\gamma$-ray light curve can be much stronger, and the dips in the $\gamma$-ray spectra reach the maximum at energies close to $\sim$30 GeV (see Fig.~\ref{fig2}c,d and~g,h). 
Note however, that the absorption effects are large in the considered above cases since we assumed very small inclination angles for the orbital planes of these stars.

On the other hand, Abramowski et al. (2009) calculated the optical depths for a few 
S type stars in the Galactic Centre applying their relatively large inclinations angles. They found that the absorption effects due to a few of these S stars are rather weak, e.g. the maximum reduction factor for S2 star $<0.99$. We have repeated their calculations for the S2 star for which the 
updated orbital parameters are reported in Abuter et al. (2019) and  physical parameters (surface temperature, stellar radius) in Habibi et al. (2017).
We confirm that for the large values of the inclination angles (Abuter et al. 2019)
the maximum optical depth due to the radiation of S2 star is $\sim10^{-4}$.
So then, the specific stars considered by Abramowski et al. (2009) are not able to 
be responsible for observable absorption effects.
Note however, that the orbits of stars within central stellar cluster are expected to be quite dynamic since these stars can suffer relatively frequent gravitational interactions during which orbital parameters of stars can change completely.

\section{Abundance of absorption events}\label{sec:abund}

We estimate how often such absorption dips can be observed in the case of the stars within Central Star Cluster around Sgr~A$^\star$. In fact, large numbers of ($\sim 10^3$) massive stars are expected to orbit around the SMBH in Sgr$^\star$ (see Introduction). 
Let us estimate the probability that any star lays on the line of sight towards the observer during the specific period of time. This probability can be estimated as the ratio of the surface covered
by the moving stars within a specific period of time to the whole surface of the sphere in the case of a homogeneous distribution of the stars on the celestial sphere. Then, the probability of detection of a dip during a period of one year ($T = 1T_{\rm yr}$~yr) can be estimated as:
\begin{eqnarray}  
P_\mathrm{1yr} = 2R_{\gamma}v_\star T N_{\rm O}/4\pi D^2
\sim  0.09M_8^{1/2}T_{\rm yr} N_{3} T_{4.5}^3 R_{12}^2/D_{-2}^{2.5},
\label{eq6} 
\end{eqnarray}
for $N_{\rm O} = 10^3N_3$ O type stars with typical parameters, radius $10^{12}$~cm and surface temperature $3\times 10^4$~K. 
In the case of Wolf-Rayet stars, the surface temperature $T_{\rm WR} = 10^5$~K and $R_{\rm WR} = 3\times 10^{11}$~cm, and luminosity $L_{\rm WR} = 1.7\times 10^6$~L$_\odot$, the coefficient in the above formula is 0.32.  

In order to investigate the effect of non-circular orbits and the distribution of the pericenter distances we perform a Monte Carlo study. 
We simulate the orbits of individual stars following the best-fitting model of \citet{2024ApJ...962...81B}.
Namely the pericenter distance is simulated from logarithmically-flat distribution in between $1$ and $10^5$~AU, while ellipticity is generated from a flat distribution with an exclusion region dependent on pericenter distance. 
We generate a number of points (100) on each simulated orbit, taking into account the varying speed of the star along the orbit. 
We then assume isotropy of the stars, thus we randomize the direction of the observer, and compute for each simulated position of the star its impact parameter. 
By counting a fraction of simulated positions with the impact parameter smaller than the $\gamma$-sphere (Eq.~\ref{eq11}), we estimate the probability that at a given random time absorption dip event is in progress. 
For the used before O star parameters this results in $p=6\times 10^{-5}  N_{3}$.
For the expected duration of each event $\sim 1$~day this corresponds to $P_\mathrm{1yr}=2.2\times 10^{-2}  N_{3}$, comparable (four times smaller) than given by simple estimates of Eq.~\ref{eq4}.

Therefore, we conclude that the dips in the $\gamma$-ray light curves at sub-TeV energies from the Sgr$^\star$ and other similar galaxies (e.g. M31) are expected to appear rather rarely. 
The O type and B type luminous stars could produce one dip every few tens of years (if the number of O type stars is $10^3$). 
In the case of galaxies with efficient star bursts (e.g. NGC~4945), which have SMBHs with a mass similar to Sgr A$^\star$ but are surrounded by super star clusters with the amount of O type stars an order of magnitude larger, dips in the $\gamma$-ray light curves are expected to also occur an order of magnitude more frequently. 
However, it should be noted that at GeV energies the measured emission of M31 is about an order of magnitude lower than that of GC \citep{2021ApJ...918..30,2017ApJ...836..208A}. 
Weaker $\gamma$-ray emission, combined with higher bulk absorption on the combined radiation field of all the stars would however hinder the detection of such events. 

\section{Detectability of absorption dips}

\begin{figure*}
    \centering
    \includegraphics[width=0.49\textwidth]{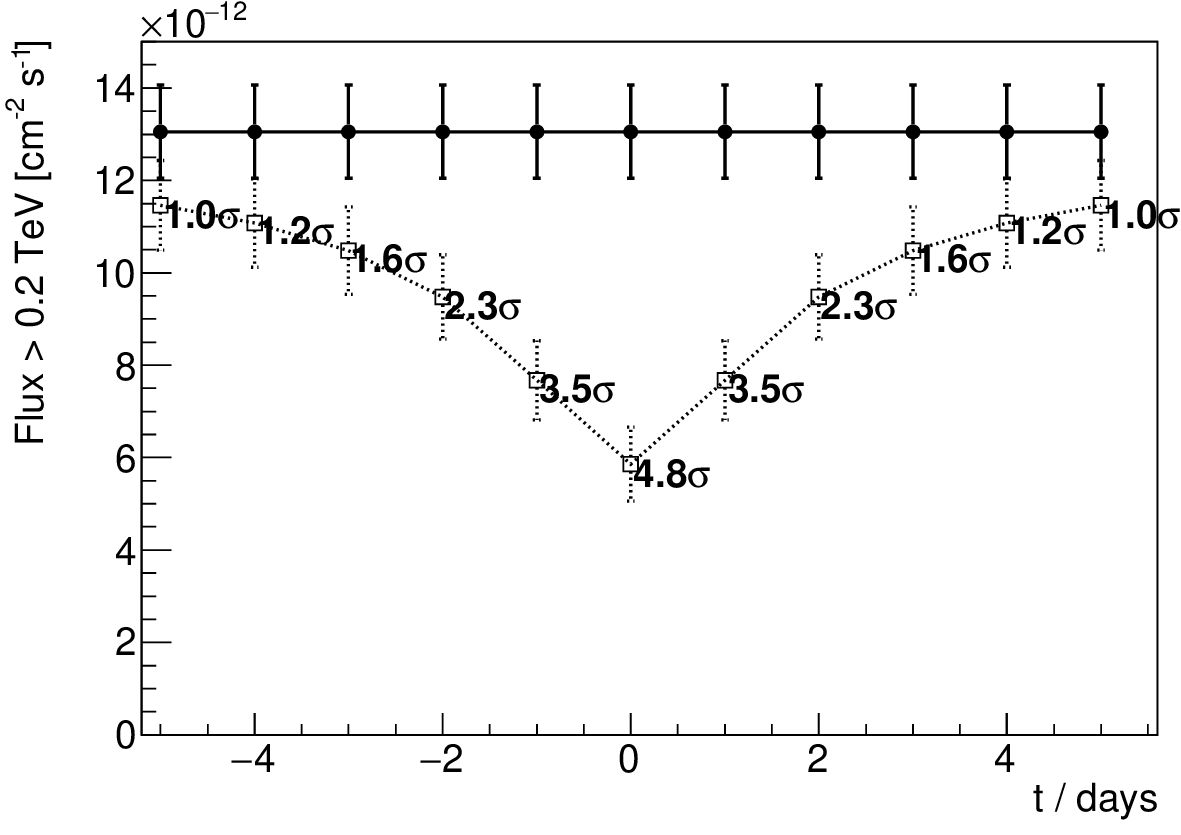}
    \includegraphics[width=0.49\textwidth]{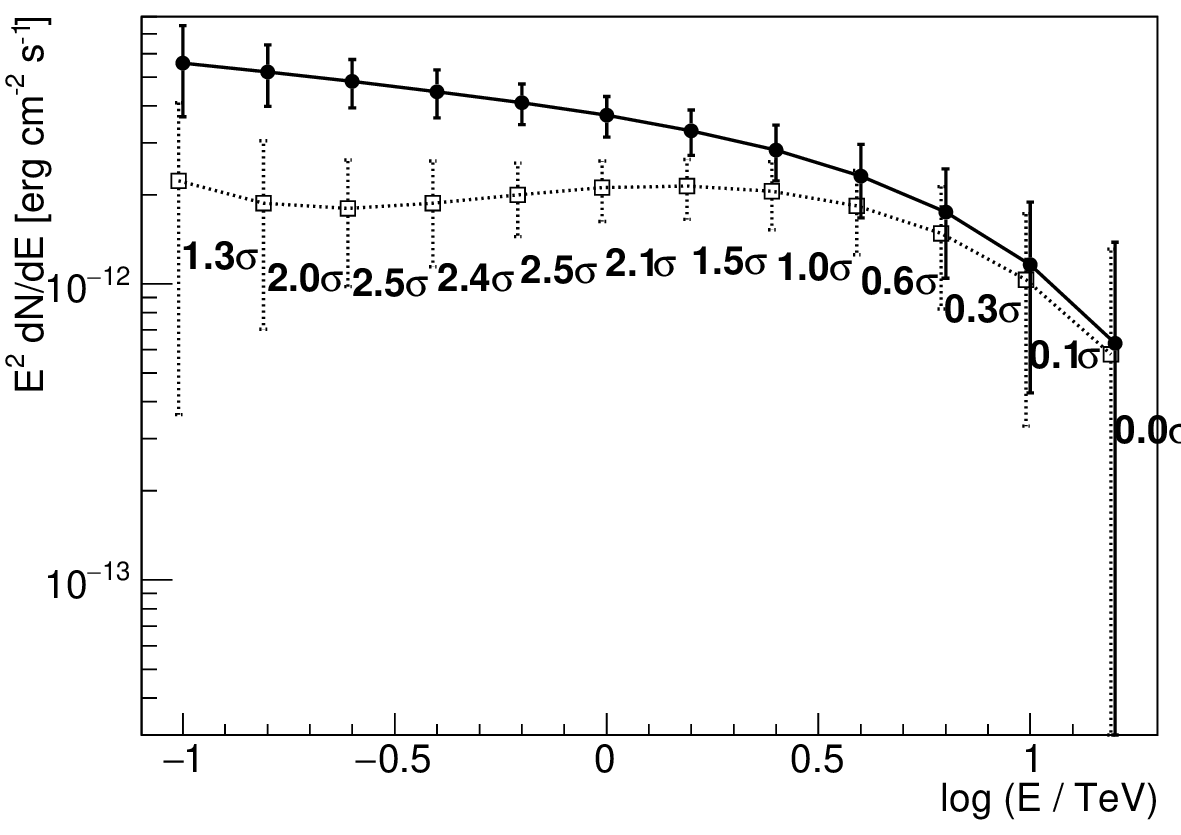}
    \caption{Left panel: Light curve above 200~GeV of the central source in the Galactic Centre as would be seen during night-by-night 3~hr-long observations by a sub-array of four LSTs without absorption (full circles and solid lines) and during the absorption (empty circles and dotted lines). 
    Next to each point a statistical significance of the difference between the two curves is reported.
    Right panel shows the corresponding spectral energy distribution during the peak of the absorption. 
    The parameters of the star are as in Fig.~\ref{fig2}a/e.}
    \label{fig3}
\end{figure*}

Next, we investigate the detectability of a single absorption dip with $\gamma$-ray ground-based and space-borne instruments.  
As explained in Section~\ref{sec:transit} in the case of WR stars the absorption can already be significant at the energies $\sim30$~GeV, accessible to \textit{Fermi}-LAT. 
The high-energy point-like Galactic Centre spectral model of \citet{2021ApJ...918..30} integrated above 30~GeV corresponds to flux of $2.7\times 10^{-10} \mathrm{cm^{-2} s^{-1}}$.
At tens of GeV, \textit{Fermi}-LAT has nearly constant acceptance of 2.5~$\mathrm {m^2 sr}$ \citep{2013arXiv1303.3514A}, therefore during regular scanning of the sky only $0.05$ events above 30~GeV are expected per day from the source. 
Thus, detection of individual absorption dips is not possible with the current generation of space $\gamma$-ray detectors. 

On the other hand, Cherenkov telescopes are much better suited for observations of the spectral features at (sub-)TeV energies.
To evaluate detectability of such a feature we take the H.E.S.S. measurement of the $\gamma$-ray emission in the central point-like source in the Galactic Centre region \citep{2016Natur.531..476H}.
The measurement is extrapolated below 200~GeV such to provide estimate of the emission above 100~GeV. 
Then we attenuate the spectrum with the reduction factor obtained in the previous section. 
We evaluate the response of the sub-array of four LST telescopes (as a part of the planned CTAO) to both spectra using the publicly available Prod5 instrument response functions (IRF, \citealp{ctao_irf}). 
As the Galactic Centre is only visible at a large zenith angle from the North location of the planned CTAO, we use the IRF corresponding to observations at the zenith angle of $60^\circ$. 
As the spectral dip is expected to be much broader than the energy resolution of the LST sub-array the energy migration effect is not expected to play an important role and for simplicity is neglected in the calculations. 
Using the collection area and residual background rates, we calculate for each energy bin the expected number of excess events ($N_{ex}$ and its uncertainty ($\Delta N_{ex}$), assuming three background control regions).
In the i-th spectral bin we compute the statistical significance of the difference as $\sigma_i = (N_{ex,0} - N_{ex,a})/\sqrt{\Delta N_{ex,0}^2+\Delta N_{ex,a}^2}$, with the $0$ and $a$ indices corresponding to the intrinsic and absorbed spectrum, respectively. 

Similarly, we also compute the expected light curve above 200~GeV by integrating the source emission with the time- and energy-dependent transmission. 
We conclude that a sub-array of 4 LSTs can achieve the detection of a dip in the integral light curve due to an O-type star during the perycenter passage, with a significance of $\sim4.8\sigma$, even in unfavourable observations at high zenith angle (see Fig.~\ref{fig3}). 
We also investigated how such an event would be seen by the Southern CTAO array (using Prod5 configuration with 14 MST and 37 SST).
In this case, the lack of LST telescopes is compensated for by observations at low zenith distance angle, resulting in a comparable threshold. The large number of telescopes in the full CTAO South array provides a similar significance of the detection in $\sim$ 1\,h. 

\section{Bulk absorption}

The absorption can occur not only on individual stars but also on the bulk radiation field of all the massive stars. 
The number density of the radiation field photons inside the central Galactic region can be estimated as:
\begin{equation}
    n_{bulk}=\frac{L_{\rm O} N_{\rm O}}{4\pi D^2 c \varepsilon} = 
    1.4\times 10^8 {{N_3}\over{D_{-2}^{2}}}~~~\mathrm{ph.~cm^{-3}},
\end{equation}
and the corresponding optical depth
\begin{equation}
    \tau_{bulk}=\sigma_{\gamma\gamma}n_{bulk}D = 0.95N_3D_{-2}^{-1}.
\end{equation}

\begin{figure*}
    \centering
    \includegraphics[width=0.49\linewidth]{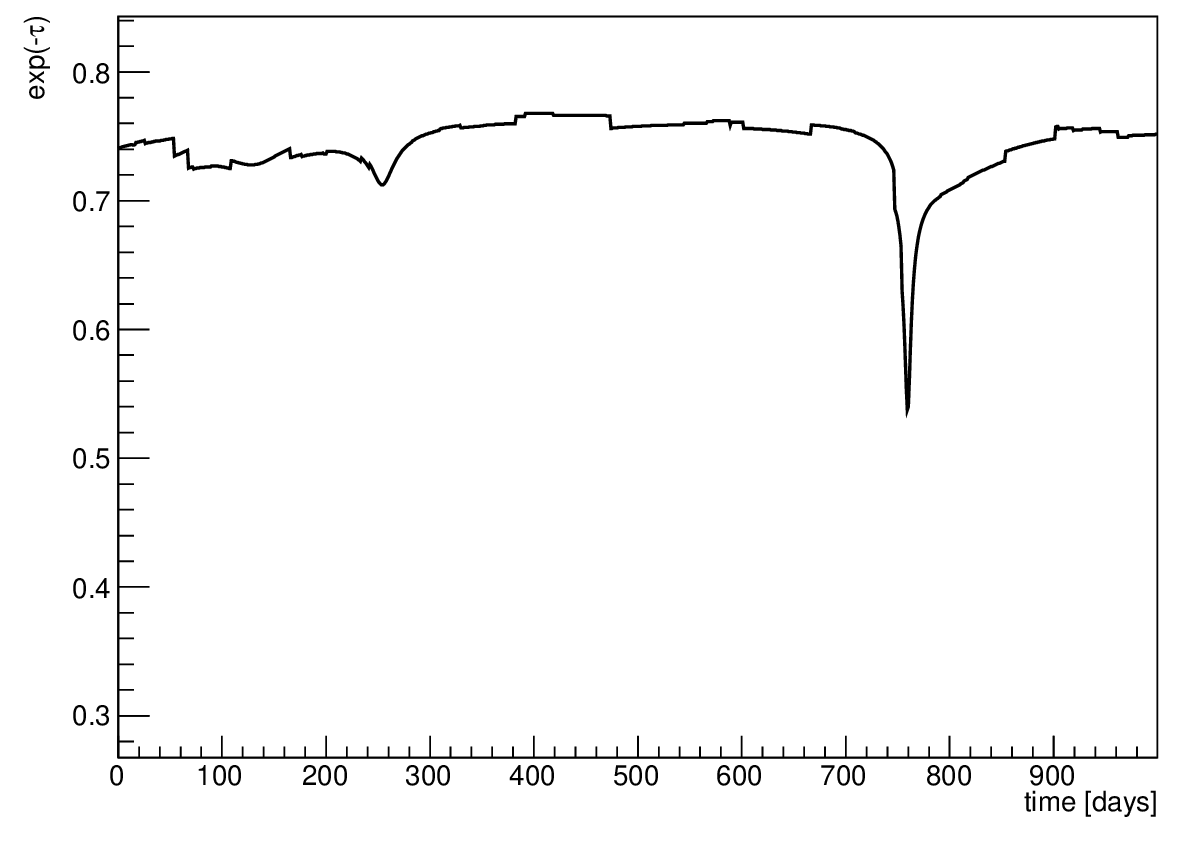}
    \includegraphics[width=0.49\linewidth]{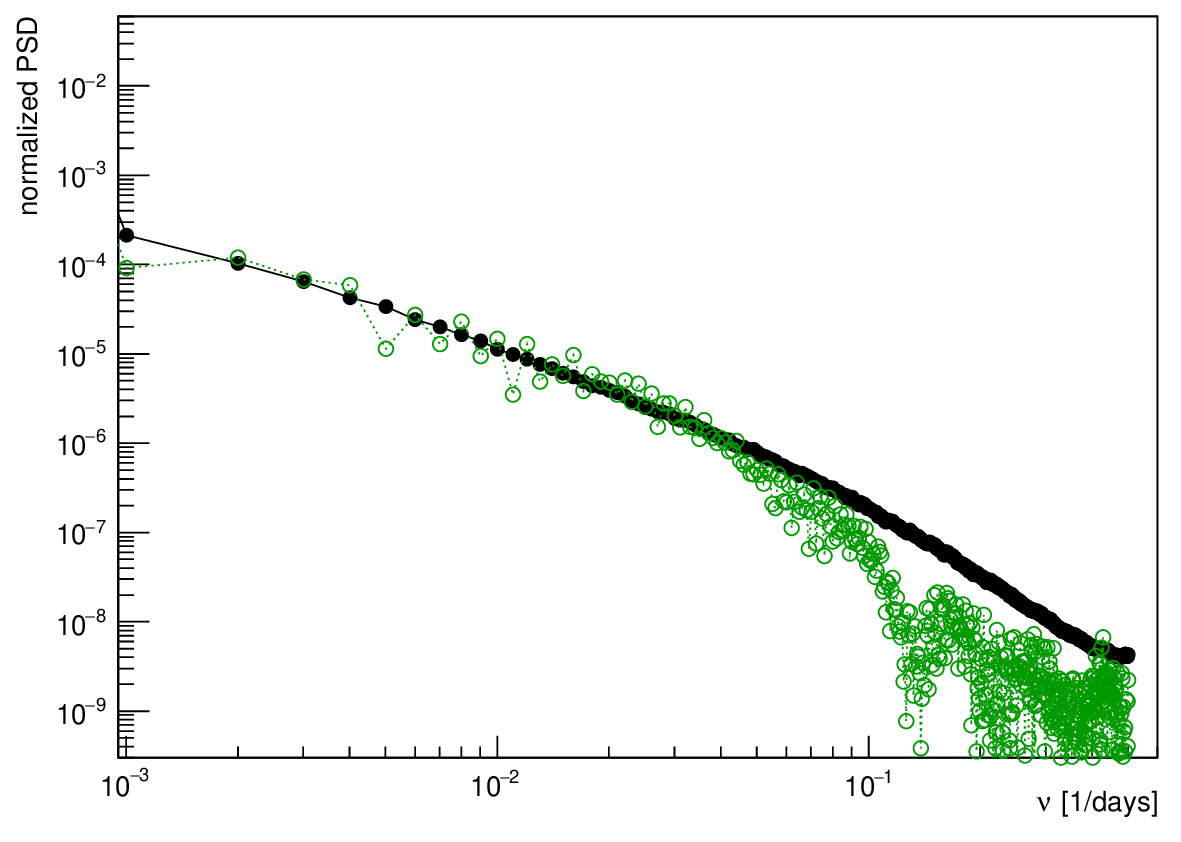}
    \caption{Effect of the combined absorption at 100 GeV by a population of 1000 O-type stars. 
    Left panel: example of a light curve (sub sample) attenuation factor  evolution due to flickering absorption on the bulk radiation field in the central Galactic region.
    Right panel: the power density spectrum of the flickering absorption for one realization of a 999-day long light curve (green empty markers) and average over all simulated light curves (black solid markers).}
    \label{fig:fft}
\end{figure*}

Since such optical depth is non-negligible, the radiation from the central Galactic region will be continuously partially absorbed, producing a flickering pattern modulated by the movement of the individual stars.
To investigate this effect, we use a model of star orbits described in Section~\ref{sec:abund} and track the Keplerian motion of the individual stars.
As a first step we assume that $10^3$ stars have the same temperature and radius typical of an O-type star. 
We then simulate the daily binned light curve of the absorbed emission at different times. 
We generate 10 realizations of such a light curve (random realizations of the assumed orbit parameter distributions), each with $10^5$ points.
The bulk absorption in this case results in an average drop by 25\% and the relative standard deviation of the flux is $\sim5\%$. 
Afterwards, we sample 999-days-long sub samples out of these light curves (see the left panel of Fig.~\ref{fig:fft} for an example light curve) and perform a Fourier transform from each and use its squared magnitude to obtain power spectral density (PSD).
Next, we average all PSDs and normalize them to the constant component (value for zero frequency).  

The resulting PSD is shown in the right panel of Fig.~\ref{fig:fft}. 
It has a roughly power-law shape with an index of $\sim-2$, typical of red noise.
This likely represents the ``random walk'' character of the absorption within the radiation field of moving stars. 
Slight steepening of the PSD with frequency is possibly due to the shape of the individual absorption events on an O-type star. 

In the next step, we validate if the mass distribution function in the central region of the Galactic Centre affects qualitatively the expected effects of the absorption.
To calculate the absorption from an arbitrary star we first consider a simple scaling relations of the temperature, $T_\star$, radius, $R_\star$, and total luminosity, $L_\star$ of the star with its mass, $M_\star$:
\begin{eqnarray}
    L_\star&=& L_O \times (M_\star/M_O)^{3.5}, \label{eqLstar}\\
    T_\star&=& T_O \times (M_\star/M_O)^{0.485}, \label{eqTstar} \\
    R_\star&=& R_O \times (M_\star/M_O)^{0.781}, \label{eqRstar}
\end{eqnarray}
with the corresponding parameters of the O-type star ($L_O=1.5\times 10^5 L_\odot$, $T_O=3\times 10^4$ K, $R_O=10^{12}$~cm). 
Eq.~\ref{eqLstar} is obtained from \citet{2005...Wiley..138}.
Eq.~\ref{eqTstar} we derive from a power-law interpolation between the O-type star and the Sun-like star.
Finally, Eq.~\ref{eqRstar} is a consequence of the former two. 
Using the computed energy-dependent optical depth $\tau_O(E)$ for a $\gamma$ ray passing an O-type star at impact $I_O$ we calculate the optical depth for an arbitrary star and impact parameter:
\begin{equation}
    \tau(E, I, L_\star, T_\star) = (L_\star/L_O)^3 (T_\star/T_O)^{-1} \tau_O(E T_\star/T_O)\times(I/I_O)^{-1}
\end{equation}

We evaluate the PSD for two mass distribution functions, a power-law with a slope of $-2.3$ (\citealp{1955ApJ...121..161, 2002Science...295..82}, the standard distribution for main sequence stars) or $-0.45$ (\citealp{2006ApJ...643..1011, 2010ApJ...708..834}, as observed in the central part of the Galactic nuclear cluster), spreading between $1 M_\odot$ and $120 M_\odot$.
The total number of stars is normalized to have $\sim 100$ stars with masses greater than $30 M_\odot$.
In the former case this sums up to $\sim9100$ stars,  of which 450 are more massive than $10 M_\odot$.
In the latter: $\sim 180$ stars in total, most of which ($\sim140$) are more massive than $10 M_\odot$. 
The average absorption caused by the scenario with the slope of $-2.3$ is comparable to the O-type-only scenario: about 25\% of the flux is absorbed, however, with a larger relative standard deviation of the distribution of fluxes: $\sim 10\%$.
The slope $-0.45$ scenario results in somewhat stronger ($\sim 45\%$ of the original flux is absorbed on average) and more variable (relative standard deviation of $\sim 20\%$) average absorption, due to a considerable number of very luminous stars.

\begin{figure*}
    \centering
    \includegraphics[width=0.49\linewidth]{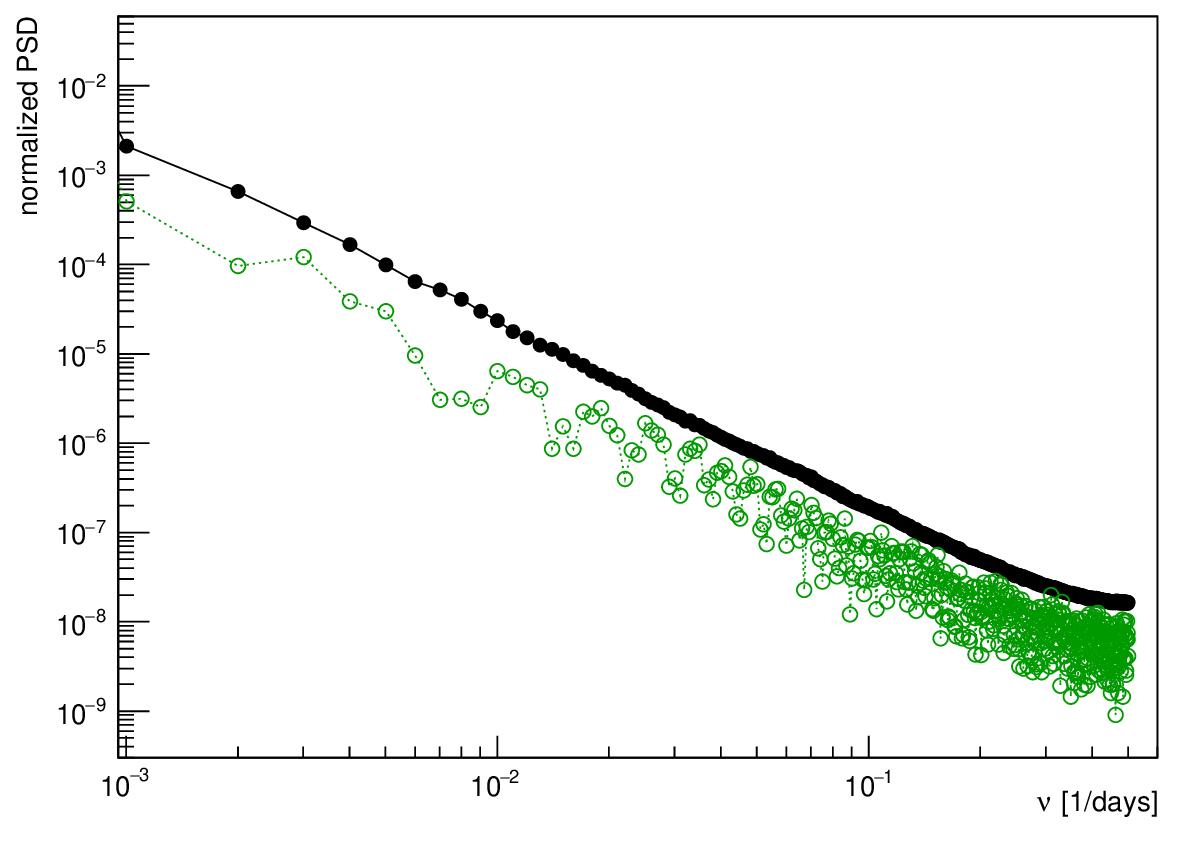}
    \includegraphics[width=0.49\linewidth]{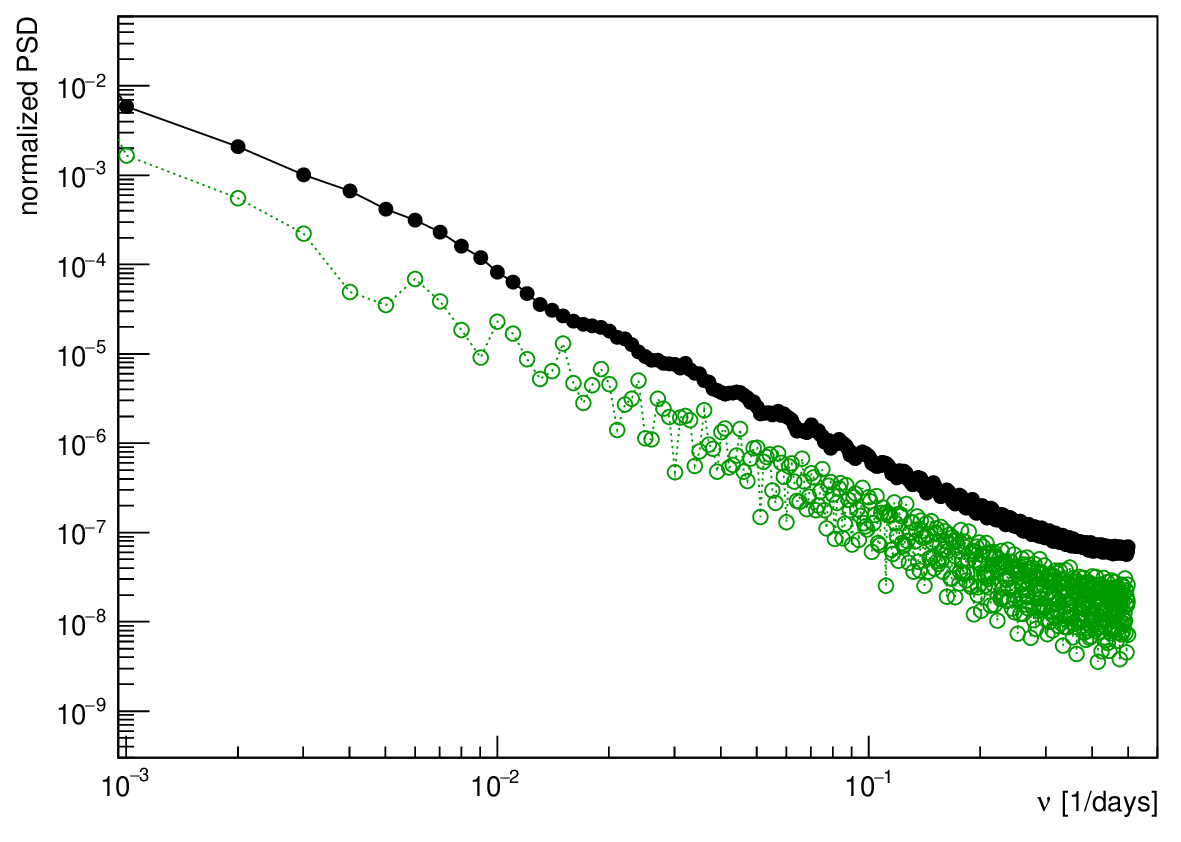}
    \caption{Normalized PSD for a distribution of star masses with index of $-2.3$ (left panel) and $-0.45$ (right panel).
    Green empty markers correspond to one realization of a 999-day long light curve  and average over all simulated light curves is shown with black solid markers.}
    \label{fig:fft_mass}
\end{figure*}

The results are shown in Fig.~\ref{fig:fft_mass}.
The resulting PSD shows similar red-noise behaviour in all the models. 
However, contrary to the O-type-only model, there is no clear steepening of the PSD above the frequency of 0.1 day$^{-1}$. 
Therefore this effect was likely caused by stacking up multiple absorption events on the same star type, and is diluted by broad distribution of the star masses. 
The stronger bulk absorption in the case of the hard mass distribution is also increasing the normalization of the PSD (computed with respect to the constant component). 

We note moreover that the absorption process of $\gamma$ rays in the nuclear star cluster (NSC) and in the vicinity of massive stars can occur in different modes. 
This is related to the magnetic fields that are stronger near the individual massive stars than the average field in the NSC. 
Therefore, the fate of the secondary $e^\pm$, pairs from the absorption process of primary $\gamma$ rays, will differ significantly. $e^\pm$ pairs lose energy mainly on the synchrotron process in the vicinity of the star, dominating over their IC energy losses.
Therefore, secondary pairs are not able to contribute significantly to the time-dependent $\gamma$ ray emission passing in the vicinity of the star.
On the other hand, in a weak magnetic field of NSC, $e^\pm$ pairs mainly interact with the soft radiation producing the next generation of $\gamma$ rays, which will partially contribute to the absorption dip in the bulk absorption case. 
In the case of the massive star we assume the surface magnetic field  $B_\star\sim 10^3$~Gs. The magnetic field strength in the central parsec within NSC is of the order of $B_{\rm NSC}\sim$2 mGs (Aitken et al. 1998). We apply a simple model for the magnetic field around a massive star, i.e. radial up to 10 r ($B(r) = B_\star r^{-2}$) and toroidal above ($B\propto r^{-1}$).   
We estimate the density of the magnetic field energy at the stellar surface on $\rho_{\rm B} = B_\star^2/8\pi\sim 
4\times 10^4$~erg~cm$^{-3}$. The energy density of the magnetic field within NSC is 
$U_{\rm B} =  B_{\rm NSC}^2/8\pi\sim 1.6\times 10^{\rm -7}$~erg~cm$^{-3}$. 
On the other hand, the energy density of radiation from the massive star is 
$\rho_{\rm ph}\sim 1.6\times 10^3/r^2$~erg~cm$^{\rm -3}$ and the energy density for photons within the NSC is $\rho\sim L_{\rm O} N_{\rm O}/(4\pi D^2 c)\sim 
1.8\times 10^{-3}N_{\rm 3}/D_{-2}^2$~erg~cm$^{-3}$. We conclude that the secondary e$^\pm$ pairs in NSC lose energy mainly on the IC process since the energy density of radiation in NSC dominates over the energy density of the magnetic field.
These e$^\pm$ pairs will contribute to bulk absorption. As a consequence, the effect of bulk absorption will be reduced.

If the secondary $e^\pm$ pairs are confined to the region with strong magnetic field, then they might be responsible for the lower energy synchrotron  X-ray
flare accompanying the close transition of the massive star. However, it is difficult to consider in detail the relative importance of the $\gamma$-ray decline and the X-ray flare since it strongly depends on the geometry of the magnetic field. The X-ray emission will likely appear in different directions than the direction to the observer. 

We also note that 96$\%$ of observed stars within central parsec are  luminous post main-sequence stars  with masses $0.5 - 4 M_\odot$,
temperatures $\sim$3500 - 3700 K and luminosities of the order of $L_\star\sim 10^3 L_\odot$ (Genzel et al.~2010). The radii of such stars would correspond to $\sim 81 R_\odot$. We estimate the thermal photon density at a distance of $r = R/R_\odot$ from the surface of such stars on $n_{\rm ph} =  2.2\times 10^{11}/r^2$ ph.~cm$^{-3}$ (see Eq.~2). The optical depth for $\gamma$ rays in this radiation field is
estimated on $\tau\sim n_{\rm ph} R \sigma_{\rm \gamma\gamma}\sim  0.28/r$. We conclude that these post-main-sequence stars can also produce significant absorption effects of the $\gamma$ rays but within the energy range close to $\sim$300 GeV, i.e. closer to the maximum sensitivity range of the future CTA. We also estimate the bulk radiation field due to this population of stars,
assuming as an example that $N = 10^4N_4$ such stars (about 6000 are observed directly, Trippe at el.~2008) are around GC on 
$n_{\rm ph} = L_\star N_\star /(4\pi D^2 c \varepsilon)\sim 8\times 10^7N_4/D_{-2}^2$~ph.~cm$^{-3}$ and the optical depth for $\gamma$ rays is
$\tau\sim n_{\bf ph} D \sigma_{\rm \gamma\gamma}\sim 0.5N_4/D_{-2}$. We conclude that for such parameters, the absorption of $\gamma$ rays on the bulk soft radiation produced by the post-main-sequence stars is comparable to the absorption of $\gamma$ rays on individual stars
and also on the bulk radiation from the O type stars.

There is also another group of interesting luminous objects in the GC the so-called G objects. They are supposed to be stars enshrouded in dense concentrations of gas and dust with the size of $\sim  100$~AU and an effective temperature of 500-600 K (with the luminosity $\sim 10^4 L_\odot$, see Zaja\'cek et al. 2014). At present about ten such objects have been observed (Ciurlo et al. 2020).
The characteristic effective temperature of these objects is about 50 times lower than the temperatures considered above for O type stars. Therefore, such transiting objects through the line of sight to the observer should result in the appearance of the absorption dip at energies close to $\sim 1.5$~TeV. 
The density of the infra-red photons close to the surface of the G objects is $n_{\rm ph} =  8.4\times 10^8/r^2$~ph.~cm$^{-3}$ and the optical depth for these $\gamma$ rays in the infra-red radiation of a specific G object should be of the order of $\sim 0.14/r$. We conclude that the infra-red radiation from specific G objects is not able to efficiently absorb TeV $\gamma$ rays.
We also calculated the optical depth for $\gamma$ rays propagating in the infra-red radiation produced by such hundred G objects ($N_{\rm G} = 100N_{2}$) around the Galactic Centre. We estimate the average density of such infra-red photons on $n_{\rm ph} = 4.5\times 10^7 N_{2}/D_{-2}^2$~ph.~cm$^{-3}$ and the optical depth for $\gamma$ rays on $\tau\sim 0.3N_{2}/D_{-2}$. We conclude that the effect of absorption of TeV $\gamma$ rays in the bulk infra-red radiation produced by G objects is rather mild, keeping in mind that the typical distance of the G objects from the Sgr~A$^\circ$ falls into the range of distances of the S stars, i.e. between 0.01 - 0.1 pc (Ciurlo et al. 2020).

\section{Discussion and conclusions}

During the last two decades, the GeV-TeV $\gamma$-ray emission has been discovered from many
nuclei of active and some ``normal'' galaxies. However, its precise localization, either in the direct vicinity of the SMBH horizon or in the intermediate-scale jet, is directly impossible due to the poor angular resolution of the present $\gamma$-ray observatories. Here, we propose a method which allows cons-training the location (and also structure) of the GeV-TeV $\gamma$-ray emission region within galactic nuclei.
In this method, we employ the effect of absorption of sub-TeV $\gamma$-ray emission, which originated close to the horizon of the SMBH in the galactic centre, in the radiation of a luminous star from a dense star cluster
around the SMBH. The $\gamma$ rays are efficiently absorbed by stellar radiation when the star passes in front of the galactic nucleus, as seen at the location of the observer. For typical parameters of the
SMBH (mass in the range between $10^6 - 10^8$~M$_\odot$) and the stellar cluster (radius $<$0.1-0.01~pc)  characteristic dip appears in the $\gamma$-ray light curve from the galactic nuclei, lasting from a fraction of a day up to a few tens of days. In the case of point-like $\gamma$-ray emission regions, the duration of the dip depends only on the parameters of the SMBH and the stellar cluster. However, if the $\gamma$-ray source has the internal structure comparable to the radius of the $\gamma$-sphere of the transiting luminous stars
(which is typically $\sim$30-100 radii of the star, i.e. $\sim (3 - 10)\times 10^{13}$~cm), then this transit method should allow putting constraints on the size and shape of the $\gamma$-ray emission region.

The dips from the absorption on the individual luminous stars will be overlaid on the global absorption in the bulk radiation field of the NSC. 
The movement of the stars in the cluster will reflect in a flickering of the observed $\gamma$-ray emission.
The power spectrum of such an observed emission is expected to follow a red noise shape.

The above considered absorption effects appear mainly in the sub-TeV part of the $\gamma$-ray spectrum, in the region between a few tens of GeV and a few TeV. This range depends on the surface temperature of the luminous stars.     

This method could be applied to the Central Star Cluster around SMBH in Sgr A$^\star$. In the case of Sgr A$^\star$, the estimated rate of stellar transits (dips in the sub-TeV $\gamma$-ray light curve) is rather low, of the order of one per ten years. However, it might be easily detected with the system of the Large-Sized Telescope (LST) of the Cherenkov Telescope Array (CTA), currently under construction.  

Clearly larger rates of $\gamma$-ray dips ($\sim$ one per few years) are expected in the case of star-burst galaxies
such as NGC 4945 which central SMBH, with the mass of the order of that in Sgr A$^\star$, is surrounded by
several star clusters (star super cluster) with the total mass of $\sim 10^7$~M$_\odot$. 
However, the $\gamma$-ray emission in this case will be affected also by a much stronger bulk absorption. 

It is interesting to consider the applicability of such an absorption effect to the Lorentz Invariance Violation searches (see \citealp{2022PrPNP.12503948A} for a recent review). 
As long as the intrinsic emission region is much smaller than the size of the $\gamma$-sphere the absorption dip is well defined and thus not dependent on the details of the source emission model. 
However, not very large distance from the Galactic Centre combined with rather long typical durations of the absorption dip would make the method only sensitive to Quantum Gravity effects with energy scales $\lesssim 10^{10}$~GeV and $\lesssim 10^7$~GeV for linear and quadratic scales, respectively.
Better limits can be achieved if the absorption dip is seen in one of the Local Group galaxies, ($\lesssim 10^{13}$~GeV and  $\lesssim 10^8$~GeV respectively); however they are still far from the Planck energy scale, and not competitive with other LIV effect search methods. 

The presence of such symmetric dip features might also be searched in the archival light curves from many extra-galactic sources that show sub-TeV $\gamma$-ray emission from the vicinity of their SMBHs.  We have systematic data covering the period of the last $\sim$20 years from few Cherenkov telescope arrays (H.E.S.S., MAGIC, VERITAS).

\begin{acknowledgements}
We thank the anonymous Referee for valuable comments and suggestions which allowed us to improve the content of the article. This research is supported by the grants from the Polish National Science Centre No. 2019/33/B/ST9/01904 (WB and MU) and No. 2023/50/A/ST9/00254 (JS).
For the purpose of Open Access, the author has applied a CC-BY public copyright licence to any Author Accepted Manuscript (AAM) version arising from this submission.
\end{acknowledgements}

\end{document}